\documentclass[iop]{emulateapj}
\usepackage{url}
\newcommand{\DMunits}{\mbox{pc cm}^{-3}}

\def\bfref{}

\begin{document}
	
\title{Fast Radio Burst Discovered in the Arecibo Pulsar ALFA Survey}
\author{
L.~G.~Spitler\altaffilmark{1},
J.~M.~Cordes\altaffilmark{2},
J.~W.~T.~Hessels\altaffilmark{3,4},
D.~R.~Lorimer\altaffilmark{5},
M.~A.~McLaughlin\altaffilmark{5},
S.~Chatterjee\altaffilmark{2},
F.~Crawford\altaffilmark{6},
J.~S.~Deneva\altaffilmark{7},
V.~M.~Kaspi\altaffilmark{8},
R.~S.~Wharton\altaffilmark{2},
B.~Allen\altaffilmark{9,10,11},
S.~Bogdanov\altaffilmark{12},
A.~Brazier\altaffilmark{2},
F.~Camilo\altaffilmark{12,13},
P.~C.~C.~Freire\altaffilmark{1},
F.~A.~Jenet\altaffilmark{14},
C.~Karako--Argaman\altaffilmark{8},
B.~Knispel\altaffilmark{10,11},
P.~Lazarus\altaffilmark{1},
K.~J.~Lee\altaffilmark{15,1},
J.~van~Leeuwen\altaffilmark{3,4}, 
R.~Lynch\altaffilmark{8},
A.~G.~Lyne\altaffilmark{16},
S.~M.~Ransom\altaffilmark{17},
P.~Scholz\altaffilmark{8},
X.~Siemens\altaffilmark{9},
I.~H.~Stairs\altaffilmark{18},
K.~Stovall\altaffilmark{19},
J.~K.~Swiggum\altaffilmark{5},
A.~Venkataraman\altaffilmark{13},
W.~W.~Zhu\altaffilmark{18},
C.~Aulbert\altaffilmark{11},
H.~Fehrmann\altaffilmark{11}
}
\altaffiltext{1}{Max-Planck-Institut f\"ur Radioastronomie, D-53121 Bonn, Germany}
\altaffiltext{2}{Department of Astronomy and Space Sciences,  Cornell University, Ithaca, NY 14853, USA}
\altaffiltext{3}{ASTRON, Netherlands Institute for Radio Astronomy, Postbus 2, 7990 AA, Dwingeloo, The Netherlands} 
\altaffiltext{4}{Astronomical Institute ``Anton Pannekoek'', University of Amsterdam, Science Park 904, 1098 XH Amsterdam, The Netherlands}
\altaffiltext{5}{Department of Physics and Astronomy, West Virginia University, Morgantown, WV 26506, USA}
\altaffiltext{6}{Department of Physics and Astronomy, Franklin and Marshall College, Lancaster, PA 17604-3003, USA} 
\altaffiltext{7}{Naval Research Laboratory, 4555 Overlook Ave SW, Washington, DC 20375, USA}
\altaffiltext{8}{Department of Physics, McGill University, Montreal, QC H3A 2T8, Canada} 
\altaffiltext{9}{Physics Department, University of Wisconsin -- Milwaukee, Milwaukee WI 53211, USA}
\altaffiltext{10}{Leibniz Universit{\"a}t, Hannover, D-30167 Hannover, Germany}
\altaffiltext{11}{Max-Planck-Institut f\"ur Gravitationsphysik, D-30167 Hannover, Germany}
\altaffiltext{12}{Columbia Astrophysics Laboratory, Columbia University,  New York, NY 10027, USA}
\altaffiltext{13}{Arecibo Observatory, HC3 Box 53995, Arecibo, PR 00612, USA}
\altaffiltext{14}{Center for Gravitational Wave Astronomy, University of Texas at Brownsville, TX 78520, USA}
\altaffiltext{15}{Kavli institute for Astronomy and Astrophysics, Peking University, Beijing 100871, P. R. China}
\altaffiltext{16}{Jodrell Bank Centre for Astrophysics, School of Physics and Astronomy, University of Manchester, Manchester, M13 9PL, UK}
\altaffiltext{17}{NRAO, Charlottesville, VA 22903, USA} 
\altaffiltext{18}{Department of Physics and Astronomy, University of British Columbia, 6224 Agricultural Road Vancouver, BC V6T 1Z1, Canada}
\altaffiltext{19}{Department of Physics and Astronomy, University of New Mexico, NM, 87131, USA}

\email{lspitler@mpifr-bonn.mpg.de}

\begin{abstract}
Recent work has exploited pulsar survey data to identify temporally isolated, 
millisecond-duration radio bursts with large dispersion measures (DMs). 
These bursts have been interpreted as arising from a population of extragalactic sources, 
in which case they would provide unprecedented opportunities for probing the
intergalactic medium; they may also be linked to new source classes.
Until now, however, all so-called fast radio bursts (FRBs) have been detected with the Parkes radio telescope and
its 13-beam receiver, casting some concern about the astrophysical
nature of these signals.  Here we present FRB~121102, 
the first FRB discovery from a geographic location other than Parkes.  
FRB~121102 was found in the Galactic anti-center region in the 1.4-GHz Pulsar ALFA survey with the Arecibo Observatory
with a DM =  $557.4\; \pm\; 3\; \DMunits$, pulse width of $3\; \pm 0.5$~ms, and no evidence of interstellar scattering.
The observed delay of the signal arrival time with frequency agrees 
precisely with the expectation of dispersion through an ionized medium. 
Despite its low Galactic latitude ($b = -0.2^{\circ}$), the burst has three times the maximum Galactic
DM expected along this particular line-of-sight,
suggesting an extragalactic origin.  
A peculiar aspect of the signal is an inverted spectrum; 
we interpret this as a consequence of being detected in a sidelobe of the ALFA receiver. 
FRB~121102's brightness, duration, and the inferred event rate 
are all consistent with the properties of the previously detected Parkes bursts.  
\end{abstract}

\section{Introduction}
\label{intro}

Radio pulsar surveys sample the sky at high time resolution and are
thus sensitive to a range of time variability and source classes.  
Over the last decade, there has been renewed interest in
expanding the purview of pulsar search pipelines, which traditionally
exploit the periodic nature of pulsars, to also search for single dispersed pulses. 
This led to the discovery of Rotating Radio Transients \citep[RRATs;][]{mll+06}, 
which are believed to be pulsars that are either
highly intermittent in their radio emission or have broad pulse-energy
distributions that make them more easy to discover using this technique \citep{wsrw06}.  
Of the now nearly 100 known RRATs\footnote{\url{http://astro.phys.wvu.edu/rratalog/}}, 
the vast majority emit multiple detectable pulses per hour of on-sky time, though a few have thus
far produced only one observed pulse \citep{bb10,bbj+11}. 
The dispersion measures (DMs) of the RRATs are all consistent with a Galactic origin,
according to the NE2001 model for Galactic electron density \citep{cl02}.

Single-pulse search methods have also discovered a new class of fast radio bursts (FRBs) in wide-field
pulsar surveys using the 13-beam, 1.4-GHz receiver at the Parkes
radio telescope \citep{lbm+07, kskl12, tsb+13}. 
Most have been found far from the Galactic plane and have DMs that are anomalously high for
those lines-of-sight. \citet{lbm+07} reported the first such burst
with a Galactic latitude of $b = -42^{\circ}$ and DM = $375$\,$\DMunits$. 
The expected DM contribution for that line-of-sight from the ionized interstellar medium (ISM) in our Galaxy 
is only $25$\,$\DMunits$ according to the NE2001 model.
The DM excess has been interpreted as coming from the ionized intergalactic medium (IGM) 
and led to the conclusion that FRBs are extragalactic. 

More recently, \citet{tsb+13} reported four FRBs
with Galactic latitudes of $|b| > 40^{\circ}$ and DMs ranging from 521~to~$1072\, \DMunits$. 
The expected Galactic DM contribution along the lines-of-sight of these bursts is $30 - 46$\,$\DMunits$, 
i.e.\ only $3-6$\,\% of the observed DM can be attributed to our Galaxy.
An additional FRB candidate was reported by \citet{kskl12} and 
is at a lower Galactic latitude ($b = -4^{\circ}$) than the other five reported Parkes FRBs. 
This source could be of Galactic origin given that the measured DM = $746$\,$\DMunits$ 
is only 1.3 times the maximum expected DM from NE2001 along this line-of-sight.
The dispersion delay of all of the published FRBs are consistent with the expected $\nu^{-2}$ dispersion law. 
Additionally, the burst reported by \citet{lbm+07} showed frequency-dependent pulse broadening
that scaled as $\nu^{-4.8 \pm 0.4}$, consistent with the expected value of $-$4.0 to $-$4.4 \citep{lr99} for scattering by the ISM.
The brightest burst reported by \citet{tsb+13} showed a clear exponential tail
 and a pulse duration that scaled as $\nu^{-4.0 \pm 0.4}$. 
This provides additional credence to the interpretation that the signal is of astrophysical origin. 

Generally, FRBs have been found in
minute to hour-long individual observations; multi-hour follow-up
observations at the same sky positions have thus far failed to find repeated bursts.
Thus, FRBs are considered a different observational phenomenon from RRATs 
based on DMs in excess of the predicted Galactic contribution  
and the fact that none of the FRBs has been seen to repeat. 
{\bfref At this point, however, we can not be certain that the bursts are non-repeating. 
Detecting an astrophysical counterpart will be an important step in determining whether we expect repeated events.}

The progenitors and physical nature of the FRBs are currently unknown.
The FRBs have brightness temperatures  well in excess of thermal emission ($T_b >10^{33}$ K) and
therefore require a coherent emission process. 
One possible source of repeating, extragalactic FRBs is extremely bright, rare Crab-like giant pulses from extragalactic pulsars, 
which repeat over much longer time scales than currently constrained.  
Proposed extragalactic sources of non-repeating, fast radio transients include
evaporating primordial black holes \citep{r77}, merging neutron stars
\citep{hl01}, collapsing supramassive neutron stars \citep{fr13}, and
superconducting cosmic strings \citep{cssv12}. 
Alternatively, \citet{lsm13} suggest a repeating, Galactic source - 
flares from nearby, magnetically active stars, in which the DM excess is due to the star's corona. 
In this scenario, additional pulses could also be observed and potentially at a different DM.
Localizing FRBs with arcsecond accuracy is technically challenging 
but will help identify potential host galaxies, or stars, and multi-wavelength counterparts.

In any pulsar survey, the vast majority of statistically
significant signals are due to man-made radio frequency interference (RFI), 
which can originate far from the telescope or be locally generated. 
RFI can also mimic some of the characteristics of short-duration astronomical signals.
Thus, care is needed when interpreting
whether a particular signal is astronomical in origin, 
and claims of a new source class require due consideration and skepticism. 
{\bfref The situation is further complicated by the discovery of ``perytons" \citep{bbe+11}.
Perytons are which are short duration radio bursts observed in pulsar surveys 
over a narrow range of DMs but have patchy spectra and are observed in many beams simultaneously. }
The fact that FRBs have so far been observed with only the Parkes telescope has raised some concern - 
even though the observed brightness distribution and event rate can explain why
other experiments have so far not detected any similar signals.
  
In this article we report the discovery of an FRB with the Arecibo
Observatory. The FRB was found as part of the Pulsar ALFA (PALFA) survey
of the Galactic plane \citep{cfl+06}.  {\bfref This detection, made with a different telescope at
a different geographic location, bolsters the
astrophysical interpretation of a phenomenon seen until now only with Parkes.}
The outline for the rest of this paper is as follows. 
In \S \ref{survey}, we describe the PALFA survey 
and the observations that led to the discovery of the new FRB. 
The burst's properties are discussed in \S \ref{burst} 
and the implied FRB event rate is  described in \S \ref{rate}. 
A discussion of the possible origin of this FRB, both astrophysical and
otherwise, is outlined in \S \ref{origin}.
In \S \ref{discussion} we discuss the implications
of our discovery for FRBs in general and present our conclusions.

\section{Observations and Analysis}
\label{survey}
\begin{figure}
\begin{center}
\includegraphics[scale=0.55]{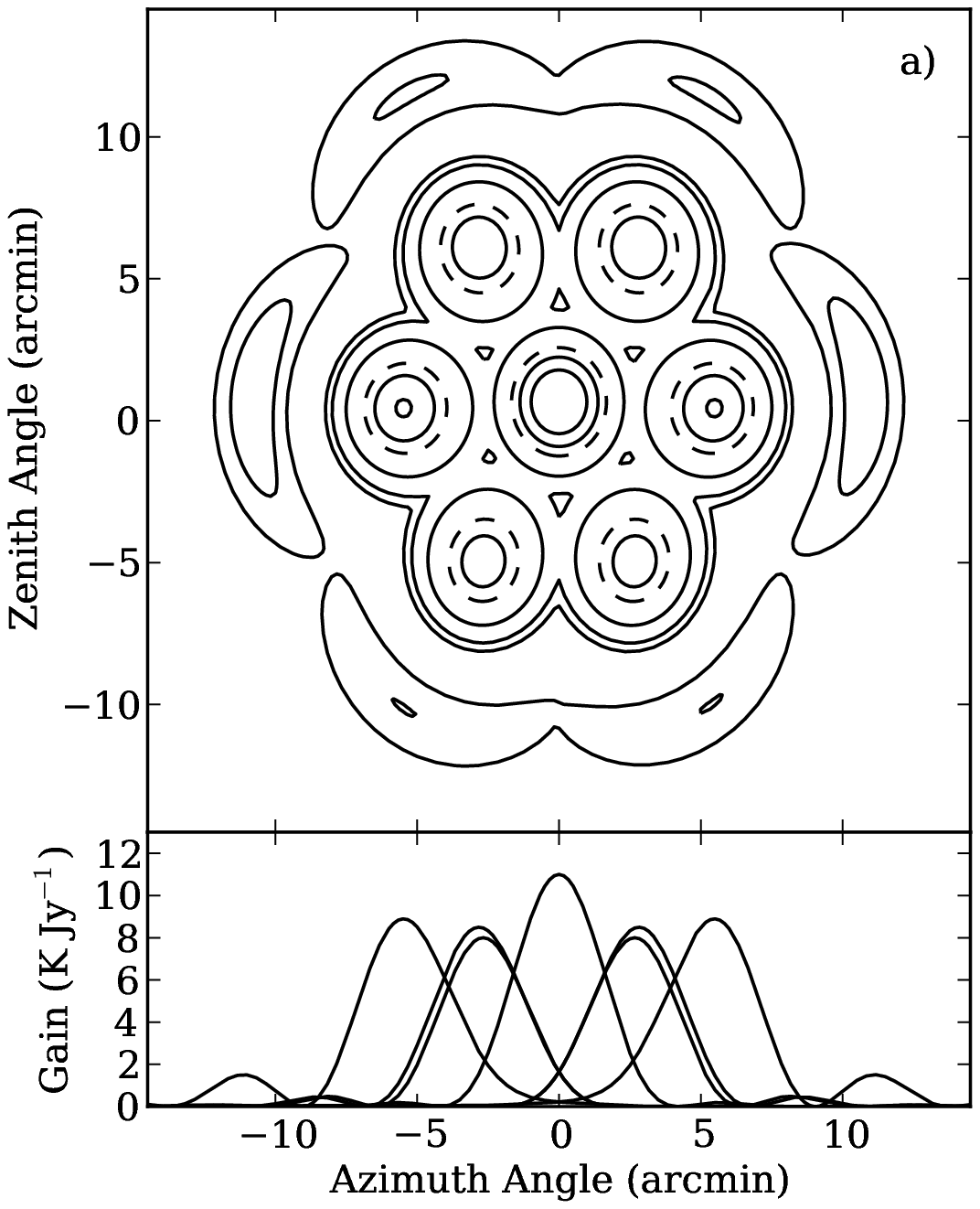}
\includegraphics[scale=0.45]{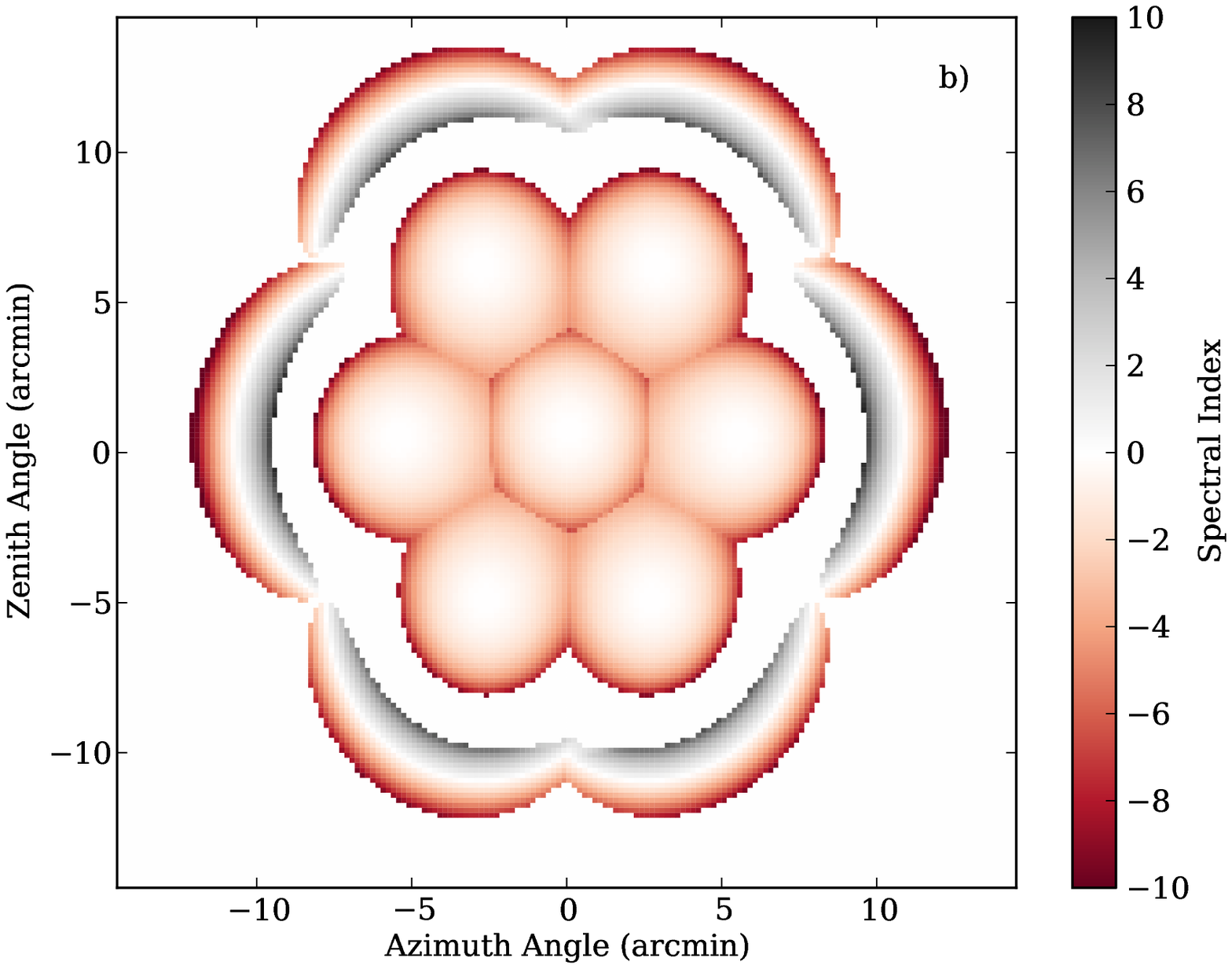}
\end{center}
\caption{Gain and spectral index maps for the ALFA receiver. 
Figure a): Contour plot of the ALFA power pattern calculated from the model described in Section~\ref{burst} at $\nu$ = 1375 MHz. The contour levels are $-$13, $-$10, $-$6, $-$3 (dashed), $-$2, and $-$1 dB (top panel). The bottom inset shows slices in azimuth for each beam, and each slice passes through the peak gain for its respective beam. Beam 1 is in the upper right, and the beam numbering proceeds clockwise. Beam 4 is, therefore, in the lower left. 
Figure b): Map of the apparent instrumental spectral index due to frequency-dependent gain variations of ALFA. The spectral indexes were calculated at the center frequencies of each subband. Only pixels with gain $>$~0.5~K~Jy$^{-1}$ were used in the calculation. The rising edge of the first sidelobe can impart a positive apparent spectral index with a magnitude that is consistent with the measured spectral index of FRB~121102. 
}
\label{fig:alfa}
\end{figure}

PALFA is a pulsar survey of the Galactic plane that uses the 305-m
Arecibo telescope and the Arecibo L-band Feed
Array\footnote{\url{http://www.naic.edu/alfa/}}  \citep[ALFA,][]{cfl+06}. ALFA is a
seven-beam feed array with a single center pixel (beam 0) surrounded
by a hexagonal ring of six pixels (beams 1---6) that
spans the frequency range 1225---1525 MHz. 
The full-width half-maximum (FWHM) of each beam is approximately 3.5$^{\prime}$, and the beams are
separated from each other by roughly one beam-width on the sky.  
(See Figure~\ref{fig:alfa} for a map of the power pattern.) 
A tessellation of three pointings is required to sample the sky to the
half-power point.  The system temperature ($T_{\rm sys}$) is 30~K. 
The on-axis gain of beam 0 is 10.4\,K\,Jy$^{-1}$, and the average
on-axis gain of the other six beams is 8.2\,K\,Jy$^{-1}$. The peak
gain of the sidelobes (about $\sim 5^{\prime}$ from the beam centers)
is 1.7\,K\,Jy$^{-1}$,  over twice as high as the on-axis gain of the Parkes beams.
The seven pixels yield an instantaneous field-of-view (FOV) of 0.022~deg$^2$ within the FWHM, 
though the effective FOV is larger as the sidelobes have enough sensitivity to detect FRBs.  

The survey began in 2004 and targets low Galactic latitudes 
($| b | \leq 5^{\circ} $) in two ranges of Galactic longitude: inner
Galaxy ($30^{\circ} < l < 78^{\circ}$) and outer Galaxy ($162^{\circ}< l < 214^{\circ}$). 
{\bfref Initially the PALFA survey data were recorded with the Wideband Arecibo Pulsar Processors \citep[WAPPs,][]{dsh00}, 
and the single-pulse analysis of these data is presented in \citet{dcm+09}. }
In March of 2009, PALFA began observing with the
Mock/PDEV\footnote{\url{http://www.naic.edu/~phil/hardware/pdev/usersGuide.pdf}}
spectrometers (hereafter, Mock spectrometers). 
{\bfref The Mock spectrometers cover the entire ALFA frequency range in 
two frequency subbands, which are separately recorded as 16-bit data. 
The 16-bit subband data are converted to 4-bit data to reduce storage requirements.
The subbands are merged prior to processing and the resulting data have 322.6\,MHz of bandwidth,
960 frequency channels, and a time resolution of 65.5\,$\mu$s. }
For a more detailed description of the Mock data see, e.g.,~\citet{l13}.


The time-frequency data are processed with a
PRESTO\footnote{\url{http://www.cv.nrao.edu/~sransom/presto/}}-based
pipeline to search for single dispersed pulses. The raw data are
cleaned of RFI using the standard PRESTO RFI excision code ({\tt rfifind}).
The raw time-frequency data are dedispersed with 5016 trial DMs
ranging from DM = 0 -- 2038\,$\DMunits$; 
the maximum DM searched is roughly twice as high as the maximum Galactic DM 
contribution expected for any line-of-sight covered in the survey.
Single pulse candidates are identified in each dedispersed time
series using a matched filtering algorithm. 
This algorithm increases the sensitivity to pulses wider than the original time
resolution of the data by convolving each dedispersed time series with
a series of boxcar matched filters \citep[for a general discussion of this technique, see][]{cm03}. 
After matched filtering, pulse candidates are identified by applying a threshold signal-to-noise ratio S/N $>$ 5.
The spectral modulation index statistic was calculated for each event to identify those caused by narrowband RFI \citep{sccs12}. 
Candidate pulses are identified by inspecting a standard set of single-pulse diagnostic plots \citep[e.g.\ Figures~5~and~6 in][]{cm03}. 
{\bfref In practice, an isolated pulse needs a S/N ratio somewhat above the threshold (S/N $\gtrsim$ 7)
in order to verify that it is astrophysical. 
A repeated source of pulses found at a S/N close to the threshold, for example from an RRAT, 
could be recognized as astrophysical because of the underlying periodicity. }

{\bfref The analysis presented here includes data recorded with the Mock spectrometers
 from beginning with their deployment in March 2009 through December 2012, 
 for a total of  $\sim$5,045 pointings in the outer Galaxy. 
 The distribution of the pointings in Galactic latitude is fairly uniform; 
 the number of pointings in one-degree bins from $|b| = 0 - 5^{\circ}$ is 8134, 5926, 7252, 7476, and 6279, respectively.
 For this analysis we do not consider the inner Galaxy pointings, 
 because of the larger DM contribution makes finding extragalactic bursts more difficult. 
 The outer Galaxy observations are conducted in piggy-back mode with our commensal partners,
 and pointing duration varies depending on their requirements.   
 Roughly 70\% of the pointings have a duration of 176 s, and 30\% have a duration of 268 seconds. 
 The total observing time of all the pointings is 283~h, or 11.8~d. }

\section{FRB 121102 Burst Description}
\label{burst}
\begin{figure}
\begin{center}
\includegraphics[scale=0.45]{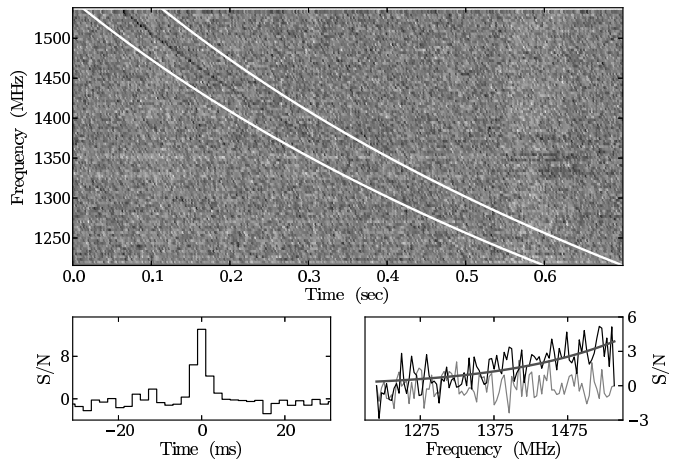}
\end{center}
\caption{Characteristic plots of FRB~121102.
In each panel the data were smoothed in time and frequency by a factor of 30 and 10, respectively. 
The top panel is a dynamic spectrum of the discovery observation showing 
the 0.7\,s during which FRB~121102 swept across the frequency band. 
The signal is seen to become significantly dimmer towards the lower part of the band, 
and some artifacts due to RFI are also visible.  
The two white curves show the expected sweep for a $\nu^{-2}$ dispersed signal at a DM = 557.4~$\DMunits$. 
The lower left panel shows the dedispersed pulse profile averaged across the bandpass. 
The lower right panel compares the on-pulse spectrum (black) with an off-pulse spectrum (light gray), 
and for reference a curve showing the fitted spectral index ($\alpha = 10$) is also overplotted (medium gray). 
The on-pulse spectrum was calculated by extracting the frequency channels 
in the dedispersed data corresponding to the peak in the pulse profile. 
The off-pulse spectrum is the extracted frequency channels for a time bin manually chosen to be far from the pulse. 
}
\label{fig:frb}
\end{figure}

A single, dispersed pulse was observed in ALFA beam 4 on 2 November 2012 (MJD = 56233.27492180) 
at 06:35:53 UT in a 176-s survey pointing toward the Galactic anticenter with an S/N = 14.  
{\bfref Because we don't know for certain where in the beam the burst occurred, 
we give the position to be the center of beam 4 ($b = -0.223^{\circ}$, $l=174.95^{\circ}$). }
The burst occurred 128~s into the observation. 
The burst properties are summarized in Table~\ref{tab:frb}. 
Following the naming convention introduced by \citet{tsb+13}, 
we henceforth refer to this event as FRB~121102. 
A frequency versus time plot of the pulse, a dedispersed pulse profile, and the spectrum of the pulse are shown in Figure~\ref{fig:frb}.  

{\bfref The DM of the pulse was calculated using a standard analysis technique also employed in pulsar timing.
The data were divided into ten frequency subbands and times-of-arrival (TOAs) were 
determined from the highest five frequency subbands. 
The lower five subbands were not included because of the low S/N. 
The best-fit DM is 557.4 $\pm$ 3.0\,$\DMunits$.
The NE2001 model predicts a maximum line-of-sight Galactic DM of 188\,$\DMunits$, 
i.e. roughly one-third of the total observed column density.  
For comparison, PSR~J0540+3207 has the larger DM of the two 
known pulsars within 5$^{\circ}$ of FRB~121102 with a DM of
62\,$\DMunits$ \citep{mhth05} and a DM-derived distance of 2.4\,kpc.

We also explored the possibility of a sweep in frequency that varies from 
the standard $\nu^{-2}$ expected from dispersion in the ionized interstellar medium, 
i.e.\  $\nu^{\beta}$, where $\beta$ is the DM index. 
The DM-fitting analysis described above determined a best-fit for the DM index of $\beta = -2.01 \pm 0.03$.
Also, a least-squares fit for deviations from the $\nu^{-2}$ law verified that a variation in $\beta$ from -2
of up to 0.05 can be tolerated given the time resolution of the data.
We also note that the value of DM is for $\beta = 2$ because the units are otherwise inappropriate.
As such, we report a final value of DM index of $\beta = -2.01 \pm 0.05$.  

The remaining burst properties were determined by a least-squares fit of the time-frequency data 
with a two-dimensional pulse model doing a grid search over a range of parameters. 
The model assumes a Gaussian pulse profile convolved with a one-sided exponential scattering tail. 
The amplitude of the Gaussian is scaled with a spectral index ($S(\nu) \propto \nu^{\alpha}$),
and the temporal location of the pulse was modeled as an absolute arrival time plus dispersive delay. 
For the least-squares fitting the DM was held constant, and the spectral index of $\tau_{\rm d}$  was fixed to be $-4.4$. 
The Gaussian FWHM pulse width, the spectral index, Gaussian amplitude, 
absolute arrival time, and pulsar broadening were all fitted. 
The Gaussian pulse width (FWHM) is 3.0 $\pm$ 0.5 ms, and we found an upper limit of $\tau_{\rm d} < 1.5$~ms at 1.4~GHz.
The residual DM smearing within a frequency channel is $0.5$\, ms and $0.9$\,ms at the top and bottom of the band, respectively.
The best-fit value was $\alpha$ = 11 but could be as low as $\alpha$ = 7. 
The fit for $\alpha$ is highly covariant with the Gaussian amplitude.
}

Every PALFA observation yields many single-pulse events that are not associated with astrophysical signals. 
A well-understood source of events is false positives from Gaussian noise.
These events are generally isolated (i.e.\ no corresponding event in neighboring trial DMs), 
have low S/Ns, and narrow temporal widths. 
RFI can also generate a large number of events, some of which mimic
the properties of astrophysical signals. Nonetheless, these can be
distinguished from astrophysical pulses in a number of ways.
{\bfref For example, RFI may peak in S/N at DM = 0\,$\DMunits$, whereas astrophysical pulses peak at a DM~$>  0\, \DMunits$. 
Although both impulsive RFI and an astrophysical pulse may span a wide range of trial DMs, the RFI will likely show no clear correlation of S/N with trial DM, while the astrophysical pulse will have a fairly symmetric reduction in S/N for trial DMs just below and above the peak value. RFI may be seen simultaneously in multiple, non-adjacent beams, while a bright, astrophysical signal may only be seen in only one beam or multiple, adjacent beams.  }
FRB~121102 exhibited all of the characteristics expected for a broadband, dispersed pulse, 
and therefore clearly stood out from all other candidate events that appeared in the pipeline output for large DMs. 

We also performed a thorough periodicity search on the discovery observation. 
We created a set of trial timeseries around the burst DM,
using an {\tt rfifind} mask to excise RFI.  Each trial DM was
searched for periodicities using PRESTO's {\tt accelsearch} and all
candidate signals with an S/N greater than 3-$\sigma$ were folded and inspected by eye.  
This closer inspection, in addition to the blind periodicity search that had 
already been done as part of the normal survey processing, also failed to reveal any periodic candidates.

Reobservations of the source position showed no additional pulses at the discovery DM.  
First, the same 7-beam ALFA survey pointing direction was repeated in a 176-s
observation on 4 November 2012. Second, targeted
follow-up was done using the L-wide single-pixel receiver and the
PUPPI spectrometer on 7 July 2013, during which the position of beam 4 was observed continuously for 1.66\,hours.
The L-wide receiver has a frequency range of $1.15
- 1.73$\,GHz and a FWHM of 3.5$^{\prime}$, while PUPPI provides a time resolution of 40.96 $\mu$s and a
bandwidth of 800 MHz, which was divided into 2048 frequency channels.
These data were processed using the single-pulse search algorithms and a narrow range of DMs spanning the burst's DM.

Lastly, because the uncertainty on the position of FRB~121102 is
larger than the full-width half maximum of the ALFA beams, we
performed a dense sampling of the region around the original beam 4
pointing center using three interleaved ALFA pointings of 2,600\,s each and
recorded with the Mock spectrometers on 9--10 December 2013.  
Combined, these covered a circular area with an approximate diameter of 17$^{\prime}$.  
An additional 2,385-s observation was conducted on 10 December 2013 using the
Arecibo 327-MHz receiver and the PUPPI backend with the telescope pointed at the center of the beam 4 from the discovery observation.
The FWHM of the 327 MHz receiver is 14$^{\prime}$, so this observation covered 
both the position of the main beam and side lobe of ALFA. 
The lower frequency observation was performed in
the event that the single bright burst seen at 1.4 GHz was part of a broad distribution
of pulses emitted from a source with a steep negative spectral index, as expected for pulsar-like coherent radiation. 
These observations were searched for single pulses over a narrow DM range
and also for periodic signals at the burst's DM. 
No additional bursts or periodic astrophysical signals were found.

One peculiar property of FRB~121102 is the observed positive
apparent spectral index of $\alpha = 7$ to 11.
{\bfref If the coherent emission process for FRBs is similar to that of pulsars, 
we would expect an intrinsically negative spectral index. 
The spectral indices of the \citet{tsb+13} are consistent with being flat. 
We therefore suspect that the observed positive spectral index is caused by
frequency-dependent gain variations of ALFA.}
To explore this possibility we developed a model for the
receiver using an asymmetric Airy function and coma lobes
but no correction for blockage from the feed support structure. 
The left panel of Figure~\ref{fig:alfa} is a map of ALFA's power pattern using this model.
We also generated a map of the induced spectral index at the center frequencies of the two Mock subbands.
The spectral index map is shown in the right panel of Figure~\ref{fig:alfa}. 
Most positions within the beam pattern, including the
entire main beam, impart a negative spectral index.  However, the
rising edge of the first sidelobe can impart a positive spectral index
bias of $0 \lesssim \alpha \lesssim 10$. We therefore conclude that the burst was 
likely detected in the sidelobe and not the main beam.  
Note that we see no evidence for the burst in the co-added dedispersed time series from pairs of
neighboring beams (beams 0 and 3 and beams 0 and 5) with S/N $>$ 4. 

It is also possible that the observed spectrum is additionally biased by RFI.  
Indeed, the lower of the two Mock subbands is significantly more affected by RFI
contamination compared with the upper subband, which may contribute
partly to the observed positive spectral index.  
In summary, given the uncertainty in the exact position, as well as the
exact beam shape and other extrinsic effects, it is unfortunately
impossible to adequately constrain the true spectral index of the
burst.

Nonetheless, the sidelobe detection hypothesis allows us to
better constrain both the position and flux of the burst. 
Conservatively, we consider a range of gain corresponding to
the inner edge of the sidelobe of 0.4 to 1.0\,K\,Jy$^{-1}$ for a mean
gain of 0.7\,K\,Jy$^{-1}$. Estimating the peak flux density ($S$) from the
radiometer equation, we have $S = 0.4_{-0.1}^{+0.4}$\,Jy, 
where we have assumed S/N = 14, $T_{\rm sys} = 30$~K, a bandwidth of 300~MHz, and a pulse duration of 3~ms.
If instead FRB~121102 was detected on axis, the flux density is $\sim$40~mJy, 
which is an order-of-magnitude weaker than any other FRB.  

\section{Event rate analysis}
\label{rate}

We can estimate the occurrence rate of FRBs from our single FRB discovery, 
the total observing time included in this analysis, and the instantaneous FOV of ALFA. 
The gain variations of the receiver (see Figure~\ref{fig:alfa})
complicate the definition of the instantaneous FOV, 
but a practical definition is the region enclosed by a minimum system gain threshold. 
We calculate the rate using two different assumptions for minimum gain. 
The first is the area enclosed by the FWHM level of the seven beams ($\Omega_{\rm FWHM}$), and 
because the sidelobes of Arecibo have comparable sensitivity to Parkes, 
we also assume a lower minimum gain that encompasses the main beams and first side lobes ($\Omega_{\rm MB+SL}$). 
We use the numerical model of the ALFA beam pattern shown in Figure~\ref{fig:alfa} 
to calculate the instantaneous FOV and a FOV-averaged system equivalent flux density $S_{\rm sys}$. 

The FWHM FOV is defined to be the area with a gain greater than half the peak gain of the outer beams, 
i.e. where G $>$ 4.1 K Jy$^{-1}$. This corresponds to $\Omega_{\rm FWHM}$ = 0.022 deg$^2$ 
and a FOV-averaged $S_{\rm sys}$ = 5 Jy. 
Using the radiometer equation, a fiducial pulse width of 1~ms, 
a bandwidth of 300~MHz, two summed polarizations, and S/N = 10, 
we get that $S_{\rm min}$ = 65 mJy. 
For 11.8 days of observing, the event rate is then $R_{S>65\; \rm mJy} = 1.6^{+6}_{-1.5} \times 10^5$~sky$^{-1}$~day$^{-1}$. 
The uncertainty interval represents the 95\% confidence interval assuming the occurrence of FRBs is Poisson distributed. 

The main beam and sidelobe FOV was defined as the region with $G > 0.4$~K~Jy$^{-1}$. 
This value of gain was chosen because it corresponds to the average 
FWHM sensitivity of the Parkes multibeam receiver. 
The instantaneous FOV is $\Omega_{\rm MB+SL}$ = 0.109 deg$^2$, 
and the FOV-averaged $S_{\rm sys}$ = 27~Jy.
Using the same parameters as above yields a minimum detectable flux density of $S_{\rm min}$ = 350 mJy.
The corresponding event rate is then $R_{S>350\; \rm mJy} = 3.1^{+12}_{-3.1} \times 10^4$~sky$^{-1}$~day$^{-1}$. 

\citet{tsb+13} have the most robust event rate of FRBs published to-date with
$R_{\rm S > 3\; Jy} = 1^{+0.6}_{-0.5} \times 10^4$~day$^{-1}$~sky$^{-1}$. 
To determine whether this rate is consistent with our inferred rates, 
one must consider the relative volumes probed given each survey's $S_{\rm min}$. 
If the FRBs do come from a population of sources at $z \gtrsim 1$, 
one must also account for cosmological effects, i.e. simple Euclidean geometry is no longer valid.
\citet{lkmj13} introduce a model for the FRB population that properly handles
the effect of cosmology on the detection rate, 
and for concreteness, they scale their predictions to the \citet{tsb+13} FRB properties. 
Scaling our event rates using this prescription, we find that our inferred rates are 
roughly consistent with the cosmological model, but we caution that the model is 
predicated on a large number of uncertain assumptions. 
In particular, the distance (or redshift) of a burst is estimated from the observed DM
and requires making an assumption about the contribution of the host Galaxy, which is highly uncertain.
Furthermore, the intrinsic emission properties of the FRBs (e.g.\ spectral index, beaming fraction) is also known.

\section{Origin of the Pulse}
\label{origin}
In this section we describe three possible origins for FRB~121102: terrestrial, Galactic, or extragalactic. 
To avoid confusion we adopt different nomenclature for the two astrophysical possibilities, 
namely FRBg and FRBx for FRBs originating from Galactic and extragalactic sources, respectively. 

\subsection{Terrestrial}
One possible terrestrial cause of FRB~121102 is RFI, 
but there are many reasons why this is unlikely.
First, the burst was seen in only a single ALFA beam. 
Strong signals due to RFI are generally seen in several
or all beams simultaneously due to their local origin \citep{bbe+11}.
To verify that this burst was localized to a single beam,
we co-added the dedispersed time
series for all beams except beam 4 and applied the same single-pulse
detection algorithms described in Section~\ref{survey}.  This resulted in
no detected pulses contemporaneous with the beam 4 signal at a S/N $> 4$.
Second, the frequency dependence of the dispersion sweep of FRB~121102
was measured to be $-2.01 \pm 0.03$, which is statistically consistent with the 
expected value of $-2$ for the propagation of a radio wave in the ISM. 
This simple relation is known to hold extremely well along Galactic lines-of-sight \citep{hsh+12}. 
Lastly, since there are no similar, isolated high-DM signals detected in our data set, 
an RFI interpretation would also require this to be quite a rare event.
 
Another possible terrestrial source are the so-called ``perytons". 
Perytons are broadband radio bursts discovered with the Parkes multibeam receiver \citep{bbe+11, kbb+12, bnm12}. 
They have typical durations of $\sim$30--50~ms (ten times longer than FRBs) and have patchy spectra. 
They are dispersed in frequency with a narrow range of timescales (typically around $\sim$360 ms)
but with dispersive frequency scalings that are not always consistent with $\nu^{-2}$. 
Most notably they are seen in many beams simultaneously
and are believed to be sidelobe detections of a bright source \citep{bbe+11} 
or near-field detections of atmospheric emission \citep{kon+14}. 
Whether the source is man-made or natural is still unclear.
While the spectrum of FRB~121102 may be reminiscent of the spectra of perytons, 
in all other regards it is quite different. 
The flux density of FRB~121102 decreases smoothly with decreasing frequency until it drops below the noise level,
and is therefore different than the patchy peryton spectra in which the the signal fades in and out across the bandpass.
The temporal widths of the perytons are at least two-times larger than FRB~121102. 
The dispersive sweeps of the perytons are at least two-times shorter than our FRB (per unit frequency). 
Perhaps most importantly, our burst was only seen in a single ALFA beam, 
while the perytons were always seen in multiple or all Parkes beams. 
We also note that we were explicitly looking for astrophysical-like signals, 
i.e.\ those that appear only in one or up to three neighboring beams.
Our apparent non-detection of perytons should not be taken as a strong statement on their existence, 
as we were not looking for them. 

\subsection{Galactic}
Because the observed DM is only three times the predicted DM from the NE2001 model (188$\, \DMunits$), 
it is conceivable that FRB~121102 is an FRBg, 
and the DM excess is caused by localized density enhancements along the line-of-sight. 
We have investigated the possibility of unmodeled gas by checking H$\alpha$ and HII survey catalogs. 
The position of FRB~121102 was mapped by the Wisconsin H$\alpha$ Mapper (WHAM)
Northern Sky Survey \citep{hrt+03} with 1-degree spatial resolution \citep[see also][]{f03}. 
The emission measure (EM) inferred from the H$\alpha$ intensity is EM$=28\; \rm pc\; cm^{-6}$. 
The NE2001 model predicts EM = 15 to 70 $\rm pc\; cm^{-6}$ using an outer scale for the electron-density spectrum of  10 to 100 pc, which is appropriate for the thick disk in the outer Galaxy.
We also searched for nearby HII regions in two complementary catalogs. 
\citet{pbd+03} compiled a catalog of 1442 HII regions from 24 previously published radio surveys, 
and \citet{abb+13} produced a catalog of over 8000 known and candidate HII regions using infrared images from the Wide-Field Infrared Survey Explorer (WISE) and archival radio data. 
The closest HII region in either catalog was greater than one degree away from the position of FRB~121102. 
In conclusion, we find no evidence for previously unmodeled dense gas along the line-of-sight 
that would explain the excess DM. 

\begin{figure}
\begin{center}
\includegraphics[scale=0.45]{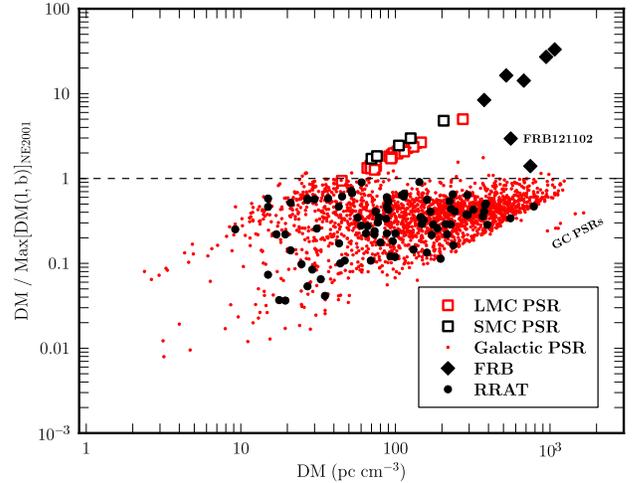}
\end{center}
\caption{DM ratio ($r_{\rm DM}$) of measured DM to maximum Galactic dispersion measure plotted against measured DM.  The maximum Galactic DM is calculated by integrating the NE2001 model to the edge of the Galaxy for each pulsar line-of-sight. The dashed line shows the maximum unity ratio expected for Galactic objects if the electron density is accurate for all lines of sight. The six pulsars near the Galactic center are clustered on the far right of the plot. The RRATs have DM ratios consistent with the rest of the Galactic pulsar population. Known pulsars in the LMC and SMC have $r_{\rm DM} \sim 1-$5, and the seven FRBs have ratios from 1.2 to 33. The \citet{kskl12} burst has the lowest DM ratio of the FRBs and is located to the lower right of FRB~121102. The \citet{lbm+07} burst and \citet{tsb+13} bursts fall along the line that extends from the LMC and SMC pulsars with the \citet{lbm+07} burst being the left-most point.  (See Section~\ref{origin} for data source references.)
}
\label{fig:dmratio}
\end{figure}

Figure~\ref{fig:dmratio} illustrates the Galactic DM excess for pulsars and FRBs
quantified by the DM ratio, $r_{\rm DM} = \rm DM_{\rm obs} / DM_{\rm NE2001, max}$, 
where $\rm DM_{obs}$ is the observed DM for a source and $\rm DM_{NE2001, max}$ 
is the DM expected for the source's line-of-sight integrated through the entire Galaxy. 
Four classes of sources are included: Galactic pulsars, RRATs, 
pulsars in the Small and Large Magellanic clouds (SMC and LMC), and FRBs.
The data for the Galactic, SMC, and LMC pulsars are from the ATNF Pulsar Catalog\footnote{\url{http://www.atnf.csiro.au/people/pulsar/psrcat/}} \citep{mhth05}. The RRAT data are from the RRATalog, and the FRB data are from this paper; \citet{lbm+07, kskl12, tsb+13}.

Galactic pulsars have $r_{\rm DM} < 1$, except for a few pulsars whose observed DM is likely 
enhanced due to unmodeled local excesses in the ISM (e.g.\ HII regions). 
{\bfref While this may suggest the DM excess of FRB~121102 could also be due to uncertainties in the NE2001 model, 
our analysis of H$\alpha$ and HII data described above makes this highly unlikely. }
The six known pulsars within $\sim$100 pc of the Galactic center (GC) are clustered at DM $\sim 1000~\DMunits$ and $r_{\rm DM} \sim 0.2 -$0.4 and are offset in DM from the rest of the pulsar population due to the increased density of the ionized ISM in the Galactic center. 
RRATs have DM ratios consistent with Galactic pulsars. 
The pulsars in the LMC and SMC have $r_{\rm DM} = 1$~to~5, 
reflecting the additional contribution of the ionized electrons in the LMC, SMC, and possibly from the local IGM. 
The FRB from \citet{kskl12} has the lowest DM ratio, which is lower even than 
some of the Galactic pulsars, suggesting this burst could be Galactic. 
In fact \citet{bm14} infer that this FRB is Galactic with a 90\% probability  
from an emission measure determined using optical spectroscopy.
The five, high-Galactic-latitude bursts from \citet{lbm+07} and \citet{tsb+13} have DM ratios greater than even the Magellanic clouds, 
which makes a Galactic interpretation difficult. 
{\bfref The LMC and SMC pulsars and high-Galactic-latitude FRBs fall long a line
because the maximum Galactic DM contribution at high Galactic latitudes is roughly constant (DM~$\sim$~50$\DMunits$).}
The DM ratio of FRB~121102 is larger than all of the Galactic pulsars but only just. We also note that the inferred extragalactic DM contribution for FRB~121102 is $\sim$~370~$\DMunits$, which is larger than for the FRB from \citet{lbm+07}. 

If the burst is an FRBg, the most likely source is an RRAT. 
RRATs have been observed with only one pulse in an epoch \citep{bb10,bbj+11}. 
But our lack of a detection in the 327 MHz follow-up observations suggests that FRB~121102
is not simply an unusually bright pulse of an otherwise weak pulsar. 
While we can not completely rule out that this burst is not an FRBg, it is highly unusual for Galactic sources. 

\subsection{Extragalactic}
The final possibility is that FRB~121102 is an extragalactic burst. 
The root of this interpretation is an observed DM in excess of the expected Galactic DM. 
The most convincing examples of FRBx's are the four \citet{tsb+13} bursts, 
and many of the properties of FRB~121102 are similar to these bursts. 
FRB~121102's observed pulse width is consistent with those
of the known population of FRBs, which have observed durations of $\sim 1
- 8$\,ms, and like FRB~121102, they generally show little to no scattering \citep{tsb+13, kskl12}.

In Section~\ref{burst} we derived a peak pulse flux density of $S = 0.4_{-0.1}^{+0.4}$\,Jy for FRB~121102. 
The flux of FRB~121102 is therefore consistent with the peak flux densities of 0.4 to 1.3 Jy 
for bursts reported by \citet{tsb+13} and 0.4 Jy reported by \citet{kskl12}. 
On the other hand, the original \citet{lbm+07} FRB~010724 had a peak flux density ($\sim$ 30 Jy)
that is still more than an order-of-magnitude larger than those discovered since.

The PALFA survey has to-date discovered only six pulsars in the outer Galaxy, 
and none of them are RRATs\footnote{\url{http://www.naic.edu/~palfa/newpulsars/}}. 
The outer Galaxy discovery with the highest DM is J0627+16 with DM = 113$\DMunits$. 
FRB~121102 is therefore much different than the other discoveries by PALFA in this region.

In summary, FRB~121102 shares many of the same observational properties with
FRBs believed to be extragalactic, and the occurrence rate is consistent with previous discoveries. 
Therefore we believe that the Arecibo FRB is also likely extragalactic. 

\section{Discussion and Conclusions}
\label{discussion}

Under the assumption that FRB~121102 is extragalactic in origin, and 
following \citet{tsb+13}, we can estimate the redshift, $z$,
of the burst based on the observed total dispersion delay across the bandpass, 
which is fortuitously for ALFA $\sim$1~ms for each 1~$\DMunits$ of DM, i.e.\ $\Delta t_{\rm obs}=552$~ms
(Figure~\ref{fig:frb}).  The contributions to $\Delta t_{\rm obs}$ are from: (i) free
electrons in the Galactic ISM ($\Delta t_{\rm ISM}$); (ii)
the intergalactic medium ($\Delta t_{\rm IGM}$); (iii) the putative host
galaxy ($\Delta t_{\rm Host}$). Adopting the NE2001 electron density for this
line-of-sight, we find $\Delta t_{\rm ISM}\simeq 184$~ms. To be consistent
with the estimates presented in \citet{tsb+13}, we make the
same assumptions about $\Delta t_{\rm IGM}$ and $\Delta t_{\rm Host}$ as presented
in their Figure~S3.  Using the DM scaling relationship for the
intergalactic medium model of \citet{i04}, we
find $\Delta t_{\rm IGM}\simeq 1200\, z$~ms. For a host galaxy DM
contribution of 100~$\DMunits$, $\Delta t_{\rm Host} \simeq 100$~ms$/(1+z)^2$. 
The condition $\Delta t_{\rm obs} = \Delta t_{\rm ISM} + \Delta t_{\rm IGM}
+ \Delta t_{\rm Host}$ is met when $z=0.26$. 
The redshift value can be taken as an upper bound because it is plausible that  
a host galaxy can contribute more to the total delay than we have assumed.
As was the case for the redshift estimates presented by \citet{tsb+13} ($z$=0.45 to 0.96), the 
contributions to $\Delta t_{\rm obs}$ are highly model dependent, 
and therefore the $z$ value should be used with caution. 

With these caveats in mind, the implied co-moving radial distance
at $z=0.26$ would be $D \sim 1$~Gpc. The FRB pseudo-luminosity
$S D^2 \sim 1 \times 10^{12}$~Jy~kpc$^2$ and 
energy output is $\sim 10^{38}$~ergs for isotropic emission and 
$\sim 10^{37}$~ergs for emission beamed over 1~steradian. 
Both values are consistent with the FRBs from  \citet{tsb+13}.
Using this estimate of the co-moving distance, we can also constrain the brightness temperature, i.e.\
$T_b \sim 1 \times 10^{34}\; D_{\rm Gpc}^2\; \rm K$, where $D_{\rm Gpc}$ is the source distance in units of Gpc.
This unphysically large brightness temperature requires a coherent emission process. 

All of the FRBs observed to date show less temporal scattering than
pulsars with similar DMs. Using a population of pulsars at low Galactic latitudes, 
\citet{bcc+04} determined an empirical relation for pulse
broadening timescale versus DM and observing frequency. 
For example, the predicted pulse broadening timescales for
$\rm DM=500$~to~$1000\; \DMunits$ are 2~to~2000\,ms at 1.4\,GHz, 
albeit with a large scatter in the observed distribution. 
By comparison, only FRB~110220 has a measurable
scattering timescale of $\sim 5$ ms with $\rm DM = 910\; \DMunits$
\citep{tsb+13}, roughly a factor of 200 less than predicted by \citet{bcc+04}. 
Using this single FRB scattering measurement, \citet{lkmj13} scale the \citet{bcc+04} relation. 
The scaled relation predicts a scattering timescale for FRB~121102 of $\sim$0.04 ms, 
which is shorter than the time resolution of the data. 
If this relation can in fact be applied broadly to FRBs, then it is not surprising that we detected no scattering. 
\citet{lkmj13} point out that for a given scattering screen, 
the largest observed scattering occurs when the screen is 
near the mid-point between the source and observer due to geometric effects. 
This suggests that the IGM or an intervening galaxy located midway along the
line-of-sight would be the most important contribution to the scattering of FRBs. 
However, \citet{cm03} show that for a source imbedded in a region of high scattering, 
for example near the center of the host galaxy or for a line-of-sight that passes through the host's galactic disk, 
the observed scattering can still be dominated by the host galaxy even at large distances.
Observations of scattering along extragalactic
lines-of-sight by \citet{lof+08} and more theoretical calculations by
\citet{mk13} suggest that scattering in the IGM is several
orders-of-magnitude lower than in the ISM, which is consistent with the 
observations of FRBs.



One caveat to the conclusions of this paper is that 
the search presented here, and all searches for dispersed radio bursts,
are optimized for signals with a $\nu^{-2}$ dispersive time delay.  
This simple approach introduces a selection effect in
what signals breach the S/N threshold used to identify candidates, 
and thus which signals are deemed worthy of close inspection.  
We note, however, that this selection effect is not severe in our case 
as the fractional bandwidth we have used here (20\%) is just
barely sufficient to see the quadratic curvature of the burst delay.


Although the poor localization of FRB~121102 prevents a detailed search for a
multi-wavelength counterpart, we searched for any major high-energy
events that were both contemporaneous and co-located on the sky.  
We checked the Gamma-ray Coordinates Network (GCN) archive of $\gamma$-ray bursts and 
found no potential association with FRB~121102.  
There are no plausible associations with X-ray transients detected by current all-sky monitors, 
and there are no observations of the field of FRB~121102 (within a 20$^{\prime}$ radius) with X-ray telescopes.
There is no source associated with this object in either the \emph{ROSAT} All Sky Survey Catalog
or the \emph{Fermi} Source Catalog.



In summary, we have described the Arecibo discovery of FRB~121102, a
single, highly dispersed pulse in the PALFA survey.  This is the first
claimed FRB detection that has been found with a telescope other than
Parkes.  The large DM excess, roughly three times what would be
expected from the Galactic ISM along this line-of-sight, the absence of
repeat bursts, and the low interstellar scattering suggest that this is an
FRB and not a Galactic emitter such as an RRAT. 
Using the occurrence rate inferred from the PALFA discovery, we predict that,
in the coming years, PALFA will find two to three more FRBs in the remaining outer Galaxy survey region. 

\acknowledgements 
We thank the referee for his or her extensive comments that significantly improved the clarity of this paper. 
We thank M.~Kramer, B.~Stappers, and R.~Ekers for useful discussions.
The Arecibo Observatory is operated by SRI International under a
cooperative agreement with the National Science Foundation
(AST-1100968), and in alliance with Ana G. M\'{e}ndez-Universidad
Metropolitana, and the Universities Space Research Association.  
These data were processed on the ATLAS cluster of the
Max-Planck-Institut f\"{u}r Gravitationsphysik/Albert-Einstein Institut, Hannover, Germany.
LGS and PCCF gratefully acknowledge financial support by the European
Research Council for the ERC Starting Grant BEACON under contract no. 279702.
JWTH acknowledges funding for this work from ERC Starting Grant DRAGNET (337062).
Work at Cornell (JMC, SC) was supported by NSF Grant 1104617.
VMK holds the Lorne Trottier Chair in Astrophysics and Cosmology
and a Canadian Research Chair in Observational Astrophysics and received additional support from 
NSERC via a Discovery Grant and Accelerator Supplement, 
by FQRNT via the Centre de Recherche Astrophysique de Qu\'ebec, 
and by the Canadian Institute for Advanced Research.
Pulsar research at UBC is supported by NSERC Discovery and 
Discovery Accelerator Supplement Grants as well as by the CFI and CANARIE.

\begin{deluxetable}{lc}
\tablecolumns{2}
\tablecaption{Observational Parameters of FRB 121102
\label{tab:frb} }
\tablehead{Parameter & Value }
\startdata
Date							&	2012 Nov 02			\\
Time							&	06:35:53 UT			\\
MJD arrival time\tablenotemark{a}	&	56233.27492180		\\
Right Ascension\tablenotemark{b}	&	05$^{\rm h}32^{\rm m}09.6^{\rm s}$	\\
Declination\tablenotemark{b}		&	33$^{\circ}05^{'}13.4^{''}$	 \\
Gal. long.\tablenotemark{b}		&	174.95$^{\circ}$		 \\
Gal. lat.\tablenotemark{b}			&	$-$0.223$^{\circ}$		 \\
DM ($\DMunits$)				&	557.4 $\pm$ 2.0     		\\
DM$_{\rm NE2001,max}$ ($\DMunits$)   &	188 					\\
Dispersion index\tablenotemark{c}		&	$-$2.01 $\pm$ 0.05		\\
Pulse width (ms)				&	3.0 $\pm$ 0.5			 \\
Pulse broadening (ms)\tablenotemark{d}			&	$<$ 1.5	\\
Flux density (Jy)\tablenotemark{e}       &       $0.4_{-0.1}^{+0.4}$         \\
Spectral index range($\alpha$) \tablenotemark{f}	&	7 to 11					
\enddata
\tablenotetext{a}{Barycentered arrival time referenced to infinite frequency.}
\tablenotetext{b}{The J2000 position of the center of beam 4.}
\tablenotetext{c}{DM $\propto \nu^{\beta}$}
\tablenotetext{d}{Flux density at 1 GHz}
\tablenotetext{e}{Flux estimation at 1.4 GHz assumes a sidelobe detection and a corresponding gain of 0.7~$\pm$~0.3~K~Jy$^{-1}$.}
\tablenotetext{f}{$S(\nu) \propto \nu^{\alpha}$}
\end{deluxetable}


\begin{thebibliography}{36}
\expandafter\ifx\csname natexlab\endcsname\relax\def\natexlab#1{#1}\fi

\bibitem[{{Anderson} {et~al.}(2013){Anderson}, {Bania}, {Balser}, {Cunningham},
  {Wenger}, {Johnstone}, \& {Armentrout}}]{abb+13}
{Anderson}, L.~D., {Bania}, T.~M., {Balser}, D.~S., {Cunningham}, V., {Wenger},
  T.~V., {Johnstone}, B.~M., \& {Armentrout}, W.~P. 2013, ArXiv e-prints

\bibitem[{{Bagchi} {et~al.}(2012){Bagchi}, {Nieves}, \& {McLaughlin}}]{bnm12}
{Bagchi}, M., {Nieves}, A.~C., \& {McLaughlin}, M. 2012, \mnras, 425, 2501

\bibitem[{{Bannister} \& {Madsen}(2014)}]{bm14}
{Bannister}, K.~W., \& {Madsen}, G.~J. 2014, ArXiv e-prints

\bibitem[{{Bhat} {et~al.}(2004){Bhat}, {Cordes}, {Camilo}, {Nice}, \&
  {Lorimer}}]{bcc+04}
{Bhat}, N.~D.~R., {Cordes}, J.~M., {Camilo}, F., {Nice}, D.~J., \& {Lorimer},
  D.~R. 2004, \apj, 605, 759

\bibitem[{{Burke-Spolaor} \& {Bailes}(2010)}]{bb10}
{Burke-Spolaor}, S., \& {Bailes}, M. 2010, \mnras, 402, 855

\bibitem[{{Burke-Spolaor} {et~al.}(2011{\natexlab{a}}){Burke-Spolaor},
  {Bailes}, {Ekers}, {Macquart}, \& {Crawford}}]{bbe+11}
{Burke-Spolaor}, S., {Bailes}, M., {Ekers}, R., {Macquart}, J.-P., \&
  {Crawford}, III, F. 2011{\natexlab{a}}, \apj, 727, 18

\bibitem[{{Burke-Spolaor} {et~al.}(2011{\natexlab{b}}){Burke-Spolaor},
  {Bailes}, {Johnston}, {Bates}, {Bhat}, {Burgay}, {D'Amico}, {Jameson},
  {Keith}, {Kramer}, {Levin}, {Milia}, {Possenti}, {Stappers}, \& {van
  Straten}}]{bbj+11}
{Burke-Spolaor}, S., {Bailes}, M., {Johnston}, S., {Bates}, S.~D., {Bhat},
  N.~D.~R., {Burgay}, M., {D'Amico}, N., {Jameson}, A., {Keith}, M.~J.,
  {Kramer}, M., {Levin}, L., {Milia}, S., {Possenti}, A., {Stappers}, B., \&
  {van Straten}, W. 2011{\natexlab{b}}, \mnras, 416, 2465

\bibitem[{{Cai} {et~al.}(2012){Cai}, {Sabancilar}, {Steer}, \&
  {Vachaspati}}]{cssv12}
{Cai}, Y.-F., {Sabancilar}, E., {Steer}, D.~A., \& {Vachaspati}, T. 2012, \prd,
  86, 043521

\bibitem[{{Cordes} {et~al.}(2006){Cordes}, {Freire}, {Lorimer}, {Camilo},
  {Champion}, {Nice}, {Ramachandran}, {Hessels}, {Vlemmings}, {van Leeuwen},
  {Ransom}, {Bhat}, {Arzoumanian}, {McLaughlin}, {Kaspi}, {Kasian}, {Deneva},
  {Reid}, {Chatterjee}, {Han}, {Backer}, {Stairs}, {Deshpande}, \&
  {Faucher-Gigu{\`e}re}}]{cfl+06}
{Cordes}, J.~M., {Freire}, P.~C.~C., {Lorimer}, D.~R., {Camilo}, F.,
  {Champion}, D.~J., {Nice}, D.~J., {Ramachandran}, R., {Hessels}, J.~W.~T.,
  {Vlemmings}, W., {van Leeuwen}, J., {Ransom}, S.~M., {Bhat}, N.~D.~R.,
  {Arzoumanian}, Z., {McLaughlin}, M.~A., {Kaspi}, V.~M., {Kasian}, L.,
  {Deneva}, J.~S., {Reid}, B., {Chatterjee}, S., {Han}, J.~L., {Backer}, D.~C.,
  {Stairs}, I.~H., {Deshpande}, A.~A., \& {Faucher-Gigu{\`e}re}, C.-A. 2006,
  \apj, 637, 446

\bibitem[{{Cordes} \& {Lazio}(2002)}]{cl02}
{Cordes}, J.~M., \& {Lazio}, T.~J.~W. 2002, ArXiv Astrophysics e-prints

\bibitem[{{Cordes} \& {McLaughlin}(2003)}]{cm03}
{Cordes}, J.~M., \& {McLaughlin}, M.~A. 2003, \apj, 596, 1142

\bibitem[{{Deneva} {et~al.}(2009){Deneva}, {Cordes}, {McLaughlin}, {Nice},
  {Lorimer}, {Crawford}, {Bhat}, {Camilo}, {Champion}, {Freire}, {Edel},
  {Kondratiev}, {Hessels}, {Jenet}, {Kasian}, {Kaspi}, {Kramer}, {Lazarus},
  {Ransom}, {Stairs}, {Stappers}, {van Leeuwen}, {Brazier}, {Venkataraman},
  {Zollweg}, \& {Bogdanov}}]{dcm+09}
{Deneva}, J.~S., {Cordes}, J.~M., {McLaughlin}, M.~A., {Nice}, D.~J.,
  {Lorimer}, D.~R., {Crawford}, F., {Bhat}, N.~D.~R., {Camilo}, F., {Champion},
  D.~J., {Freire}, P.~C.~C., {Edel}, S., {Kondratiev}, V.~I., {Hessels},
  J.~W.~T., {Jenet}, F.~A., {Kasian}, L., {Kaspi}, V.~M., {Kramer}, M.,
  {Lazarus}, P., {Ransom}, S.~M., {Stairs}, I.~H., {Stappers}, B.~W., {van
  Leeuwen}, J., {Brazier}, A., {Venkataraman}, A., {Zollweg}, J.~A., \&
  {Bogdanov}, S. 2009, \apj, 703, 2259

\bibitem[{{Dowd} {et~al.}(2000){Dowd}, {Sisk}, \& {Hagen}}]{dsh00}
{Dowd}, A., {Sisk}, W., \& {Hagen}, J. 2000, in Astronomical Society of the
  Pacific Conference Series, Vol. 202, IAU Colloq. 177: Pulsar Astronomy - 2000
  and Beyond, ed. M.~{Kramer}, N.~{Wex}, \& R.~{Wielebinski}, 275

\bibitem[{{Falcke} \& {Rezzolla}(2013)}]{fr13}
{Falcke}, H., \& {Rezzolla}, L. 2013, ArXiv e-prints

\bibitem[{{Finkbeiner}(2003)}]{f03}
{Finkbeiner}, D.~P. 2003, \apjs, 146, 407

\bibitem[{{Haffner} {et~al.}(2003){Haffner}, {Reynolds}, {Tufte}, {Madsen},
  {Jaehnig}, \& {Percival}}]{hrt+03}
{Haffner}, L.~M., {Reynolds}, R.~J., {Tufte}, S.~L., {Madsen}, G.~J.,
  {Jaehnig}, K.~P., \& {Percival}, J.~W. 2003, \apjs, 149, 405

\bibitem[{{Hansen} \& {Lyutikov}(2001)}]{hl01}
{Hansen}, B.~M.~S., \& {Lyutikov}, M. 2001, \mnras, 322, 695

\bibitem[{{Hassall} {et~al.}(2012){Hassall}, {Stappers}, {Hessels}, {Kramer},
  {Alexov}, {Anderson}, {Coenen}, {Karastergiou}, {Keane}, {Kondratiev},
  {Lazaridis}, {van Leeuwen}, {Noutsos}, {Serylak}, {Sobey}, {Verbiest},
  {Weltevrede}, {Zagkouris}, {Fender}, {Wijers}, {B{\"a}hren}, {Bell},
  {Broderick}, {Corbel}, {Daw}, {Dhillon}, {Eisl{\"o}ffel}, {Falcke},
  {Grie{\ss}meier}, {Jonker}, {Law}, {Markoff}, {Miller-Jones}, {Osten}, {Rol},
  {Scaife}, {Scheers}, {Schellart}, {Spreeuw}, {Swinbank}, {ter Veen}, {Wise},
  {Wijnands}, {Wucknitz}, {Zarka}, {Asgekar}, {Bell}, {Bentum}, {Bernardi},
  {Best}, {Bonafede}, {Boonstra}, {Brentjens}, {Brouw}, {Br{\"u}ggen},
  {Butcher}, {Ciardi}, {Garrett}, {Gerbers}, {Gunst}, {van Haarlem}, {Heald},
  {Hoeft}, {Holties}, {de Jong}, {Koopmans}, {Kuniyoshi}, {Kuper}, {Loose},
  {Maat}, {Masters}, {McKean}, {Meulman}, {Mevius}, {Munk}, {Noordam},
  {Orr{\'u}}, {Paas}, {Pandey-Pommier}, {Pandey}, {Pizzo}, {Polatidis},
  {Reich}, {R{\"o}ttgering}, {Sluman}, {Steinmetz}, {Sterks}, {Tagger}, {Tang},
  {Tasse}, {Vermeulen}, {van Weeren}, {Wijnholds}, \& {Yatawatta}}]{hsh+12}
{Hassall}, T.~E., {Stappers}, B.~W., {Hessels}, J.~W.~T., {Kramer}, M.,
  {Alexov}, A., {Anderson}, K., {Coenen}, T., {Karastergiou}, A., {Keane},
  E.~F., {Kondratiev}, V.~I., {Lazaridis}, K., {van Leeuwen}, J., {Noutsos},
  A., {Serylak}, M., {Sobey}, C., {Verbiest}, J.~P.~W., {Weltevrede}, P.,
  {Zagkouris}, K., {Fender}, R., {Wijers}, R.~A.~M.~J., {B{\"a}hren}, L.,
  {Bell}, M.~E., {Broderick}, J.~W., {Corbel}, S., {Daw}, E.~J., {Dhillon},
  V.~S., {Eisl{\"o}ffel}, J., {Falcke}, H., {Grie{\ss}meier}, J.-M., {Jonker},
  P., {Law}, C., {Markoff}, S., {Miller-Jones}, J.~C.~A., {Osten}, R., {Rol},
  E., {Scaife}, A.~M.~M., {Scheers}, B., {Schellart}, P., {Spreeuw}, H.,
  {Swinbank}, J., {ter Veen}, S., {Wise}, M.~W., {Wijnands}, R., {Wucknitz},
  O., {Zarka}, P., {Asgekar}, A., {Bell}, M.~R., {Bentum}, M.~J., {Bernardi},
  G., {Best}, P., {Bonafede}, A., {Boonstra}, A.~J., {Brentjens}, M., {Brouw},
  W.~N., {Br{\"u}ggen}, M., {Butcher}, H.~R., {Ciardi}, B., {Garrett}, M.~A.,
  {Gerbers}, M., {Gunst}, A.~W., {van Haarlem}, M.~P., {Heald}, G., {Hoeft},
  M., {Holties}, H., {de Jong}, A., {Koopmans}, L.~V.~E., {Kuniyoshi}, M.,
  {Kuper}, G., {Loose}, G.~M., {Maat}, P., {Masters}, J., {McKean}, J.~P.,
  {Meulman}, H., {Mevius}, M., {Munk}, H., {Noordam}, J.~E., {Orr{\'u}}, E.,
  {Paas}, H., {Pandey-Pommier}, M., {Pandey}, V.~N., {Pizzo}, R., {Polatidis},
  A., {Reich}, W., {R{\"o}ttgering}, H., {Sluman}, J., {Steinmetz}, M.,
  {Sterks}, C.~G.~M., {Tagger}, M., {Tang}, Y., {Tasse}, C., {Vermeulen}, R.,
  {van Weeren}, R.~J., {Wijnholds}, S.~J., \& {Yatawatta}, S. 2012, \aap, 543,
  A66

\bibitem[{{Inoue}(2004)}]{i04}
{Inoue}, S. 2004, \mnras, 348, 999

\bibitem[{{Keane} {et~al.}(2012){Keane}, {Stappers}, {Kramer}, \&
  {Lyne}}]{kskl12}
{Keane}, E.~F., {Stappers}, B.~W., {Kramer}, M., \& {Lyne}, A.~G. 2012, \mnras,
  425, L71

\bibitem[{{Kocz} {et~al.}(2012){Kocz}, {Bailes}, {Barnes}, {Burke-Spolaor}, \&
  {Levin}}]{kbb+12}
{Kocz}, J., {Bailes}, M., {Barnes}, D., {Burke-Spolaor}, S., \& {Levin}, L.
  2012, \mnras, 420, 271

\bibitem[{{Kulkarni} {et~al.}(2014){Kulkarni}, {Ofek}, {Neill}, {Zheng}, \&
  {Juric}}]{kon+14}
{Kulkarni}, S.~R., {Ofek}, E.~O., {Neill}, J.~D., {Zheng}, Z., \& {Juric}, M.
  2014, ArXiv e-prints

\bibitem[{{Lambert} \& {Rickett}(1999)}]{lr99}
{Lambert}, H.~C., \& {Rickett}, B.~J. 1999, \apj, 517, 299

\bibitem[{{Lazarus}(2013)}]{l13}
{Lazarus}, P. 2013, in IAU Symposium, Vol. 291, IAU Symposium, 35--40

\bibitem[{{Lazio} {et~al.}(2008){Lazio}, {Ojha}, {Fey}, {Kedziora-Chudczer},
  {Cordes}, {Jauncey}, \& {Lovell}}]{lof+08}
{Lazio}, T.~J.~W., {Ojha}, R., {Fey}, A.~L., {Kedziora-Chudczer}, L., {Cordes},
  J.~M., {Jauncey}, D.~L., \& {Lovell}, J.~E.~J. 2008, \apj, 672, 115

\bibitem[{{Loeb} {et~al.}(2013){Loeb}, {Shvartzvald}, \& {Maoz}}]{lsm13}
{Loeb}, A., {Shvartzvald}, Y., \& {Maoz}, D. 2013, ArXiv e-prints

\bibitem[{{Lorimer} {et~al.}(2007){Lorimer}, {Bailes}, {McLaughlin},
  {Narkevic}, \& {Crawford}}]{lbm+07}
{Lorimer}, D.~R., {Bailes}, M., {McLaughlin}, M.~A., {Narkevic}, D.~J., \&
  {Crawford}, F. 2007, Science, 318, 777

\bibitem[{{Lorimer} {et~al.}(2013){Lorimer}, {Karastergiou}, {McLaughlin}, \&
  {Johnston}}]{lkmj13}
{Lorimer}, D.~R., {Karastergiou}, A., {McLaughlin}, M.~A., \& {Johnston}, S.
  2013, \mnras, 436, L5

\bibitem[{{Macquart} \& {Koay}(2013)}]{mk13}
{Macquart}, J.-P., \& {Koay}, J.~Y. 2013, \apj, 776, 125

\bibitem[{{Manchester} {et~al.}(2005){Manchester}, {Hobbs}, {Teoh}, \&
  {Hobbs}}]{mhth05}
{Manchester}, R.~N., {Hobbs}, G.~B., {Teoh}, A., \& {Hobbs}, M. 2005, \aj, 129,
  1993

\bibitem[{{McLaughlin} {et~al.}(2006){McLaughlin}, {Lyne}, {Lorimer}, {Kramer},
  {Faulkner}, {Manchester}, {Cordes}, {Camilo}, {Possenti}, {Stairs}, {Hobbs},
  {D'Amico}, {Burgay}, \& {O'Brien}}]{mll+06}
{McLaughlin}, M.~A., {Lyne}, A.~G., {Lorimer}, D.~R., {Kramer}, M., {Faulkner},
  A.~J., {Manchester}, R.~N., {Cordes}, J.~M., {Camilo}, F., {Possenti}, A.,
  {Stairs}, I.~H., {Hobbs}, G., {D'Amico}, N., {Burgay}, M., \& {O'Brien},
  J.~T. 2006, \nat, 439, 817

\bibitem[{{Paladini} {et~al.}(2003){Paladini}, {Burigana}, {Davies}, {Maino},
  {Bersanelli}, {Cappellini}, {Platania}, \& {Smoot}}]{pbd+03}
{Paladini}, R., {Burigana}, C., {Davies}, R.~D., {Maino}, D., {Bersanelli}, M.,
  {Cappellini}, B., {Platania}, P., \& {Smoot}, G. 2003, \aap, 397, 213

\bibitem[{{Rees}(1977)}]{r77}
{Rees}, M.~J. 1977, \nat, 266, 333

\bibitem[{{Spitler} {et~al.}(2012){Spitler}, {Cordes}, {Chatterjee}, \&
  {Stone}}]{sccs12}
{Spitler}, L.~G., {Cordes}, J.~M., {Chatterjee}, S., \& {Stone}, J. 2012, \apj,
  748, 73

\bibitem[{{Thornton} {et~al.}(2013){Thornton}, {Stappers}, {Bailes},
  {Barsdell}, {Bates}, {Bhat}, {Burgay}, {Burke-Spolaor}, {Champion}, {Coster},
  {D'Amico}, {Jameson}, {Johnston}, {Keith}, {Kramer}, {Levin}, {Milia}, {Ng},
  {Possenti}, \& {van Straten}}]{tsb+13}
{Thornton}, D., {Stappers}, B., {Bailes}, M., {Barsdell}, B., {Bates}, S.,
  {Bhat}, N.~D.~R., {Burgay}, M., {Burke-Spolaor}, S., {Champion}, D.~J.,
  {Coster}, P., {D'Amico}, N., {Jameson}, A., {Johnston}, S., {Keith}, M.,
  {Kramer}, M., {Levin}, L., {Milia}, S., {Ng}, C., {Possenti}, A., \& {van
  Straten}, W. 2013, Science, 341, 53

\bibitem[{{Weltevrede} {et~al.}(2006){Weltevrede}, {Stappers}, {Rankin}, \&
  {Wright}}]{wsrw06}
{Weltevrede}, P., {Stappers}, B.~W., {Rankin}, J.~M., \& {Wright}, G.~A.~E.
  2006, \apjl, 645, L149

\end{thebibliography}

\end{document}